\documentclass[twocolumn,preprintnumbers,endnote,nofootinbib,prd]{revtex4}
\usepackage{graphicx}
\usepackage{amsmath}
\usepackage{color}

\newcommand{\beq}{\begin{equation}}
\newcommand{\eeq}{\end{equation}}
\newcommand{\bea}{\begin{eqnarray}}
\newcommand{\eea}{\end{eqnarray}}

\begin{document}

\pagestyle{plain}

\title{\boldmath An alternative interpretation for cosmic ray peaks}

\author{Doojin Kim
\footnote{email: immworry@ufl.edu}
}
\affiliation{Physics Department, University of Florida, Gainesville, FL 32611, USA}

\author{Jong-Chul Park
\footnote{email: log1079@gmail.com}
}
\affiliation{Department of Physics, Sungkyunkwan University, Suwon 440-746, Korea\\
Department of Physics, Chungnam National University, Daejeon 305-764, Korea}


\preprint{CETUP2015-019}

\begin{abstract}

We propose an {\it alternative} mechanism based upon dark matter (DM) interpretation for anomalous {\it peak} signatures in cosmic ray measurements, assuming an extended dark sector with two DM species.
This is contrasted with previous effort to explain various line-like cosmic-ray excesses in the context of DM models where the relevant DM candidate directly annihilates into Standard Model (SM) particles.
The heavier DM is assumed to annihilate to an on-shell intermediate state.
As the simplest choice, it decays directly into the lighter DM along with an unstable particle which in turn decays to a {\it pair} of SM states corresponding to the interesting cosmic anomaly.
We show that a {\it sharp continuum} energy peak can be readily generated under the proposed DM scenario, depending on dark sector particle mass spectra.
Remarkably, such a peak is {\it robustly} identified as half the mass of the unstable particle.
Furthermore, other underlying mass parameters are {\it analytically} related to the shape of energy spectrum.
We apply this idea to the two well-known line excesses in the cosmic photon spectrum: 130 GeV $\gamma$-ray line and 3.5 keV X-ray line.
Each observed peak spectrum is well-reproduced by theoretical expectation predicated upon our suggested mechanism, and moreover, our resulting best fits provide rather improved $\chi^2$ values.

\end{abstract}
\maketitle

\section{Introduction \label{sec:intro}}

Dark matter (DM) is a necessary component in the time evolution of our Universe, and its existence has been supported by a tremendous amount of astrophysical and cosmological evidence for which the relevant observations are made mostly based upon the gravitational interaction of DM (see Ref.~\cite{Bertone:2004pz} for a general review).
In fact, none of the Standard Model (SM) particles can explain various DM-related phenomena, so that the detection of any DM candidates can be not only exciting {\it per se.} but a strong sign of new physics framework beyond the Standard Model (BSM).
A great deal of experimental effort to detect DM signals has been made in three different directions such as i) direct detection experiments by observing recoil energy of target nuclei scattered off by DM, ii) indirect detection experiments by observing cosmic rays originating from DM annihilation or decay, and iii) collider searches (for example, at the Large Hardon Collider at CERN) by actively producing DM candidates and exploiting their collider signatures.
These three avenues to DM detection are complementary to one another, and have set the bounds of the viable DM mass and the associated cross section (see Ref.~\cite{Agashe:2014kda} and references therein for review).\footnote{Recently, Ref.~\cite{Davoudiasl:2015vba} proposed a general scenario, dubbed ``Inflatable DM models'', within the context of which many well-motivated DM models having too large production of DM can be remedied, hence evade the bounds without tuning of underlying parameters.}

Of those experimental attempts, indirect detection experiments have received particular attention as many of them have reported anomalous observations potentially signalling the presence of DM candidates at the locus of cosmic ray sources.
For instance, PAMELA~\cite{Adriani:2008zr}, Fermi-LAT~\cite{FermiLAT:2011ab}, and AMS-02~\cite{Aguilar:2013qda} found quite a marked rise of the positron fraction in the energy range from roughly 10 to 200 GeV, and similarly ATIC~\cite{Chang:2008aa}, FERMI-LAT~\cite{Abdo:2009zk}, and HESS~\cite{Aharonian:2009ah} reported an excess in the positron-electron combined energy spectrum between 100 and 1000 GeV.
Several photon channels showed intriguing excesses such as 3.5 keV line~\cite{Bulbul:2014sua,Boyarsky:2014jta}, 511 keV line~\cite{Jean:2003ci}, GeV bump~\cite{Goodenough:2009gk}, and 130 GeV line~\cite{Bringmann:2012vr,Weniger:2012tx}.
The positron excesses and the Galactic Center (GC) GeV $\gamma$-ray excess are featured by a continuum bump, while the other three X/$\gamma$-ray excesses showed a sharp peak within a very narrow energy range.

The latter class of excesses are particularly interesting because they can be readily connected to the DM interpretation.
As typical DM candidates behave non-relativistically, the photon energy from a DM pair annihilation (or 2-body decay) is monochromatic, being the same as (half) the DM mass.\footnote{The 511 keV $\gamma$-ray peak comes from the positronium decay.
Thus, for the explanation of the 511 keV line excess, the required is a new source of positrons which can be DM annihilation or decay.}
In this context, many DM models to address those excesses have been introduced and studied in literature: for example, Ref.~\cite{Huh:2007zw} for 511 keV line, Refs.~\cite{Kyae:2012vi, Park:2012xq} for 130 GeV line, and Ref.~\cite{Kong:2014gea} for 3.5 keV line.
In reality, the relevant signal spectrum does {\it not} appear as a $\delta$-function-like peak but is smeared to some extent because of imperfection in cosmic ray detectors.
With the assumption of a Gaussian smearing, the resultant $\gamma$-ray energy spectrum (typically) becomes {\it symmetric} with respect to the nominal peak.

This broadening effect has motivated the possibility of non-minimal DM scenarios for interpreting the narrow width of peaks as the physical, not the one induced from instrumental uncertainties.
The next-to-minimal DM models hypothesize the situation where DM particles annihilate or decay into on-shell intermediate particles that decay into photons
\cite{Ibarra:2012dw, Boddy:2015efa}.
As such an on-shell intermediate particle comes along with a fixed boost, the photon energy spectrum is characterized by a rectangular shape \cite{Ibarra:2012dw, Boddy:2015efa}.
If the mass gap between the DM and the intermediate particle is sufficiently small, hence so is the boost factor, then the photon energy spectrum becomes narrow enough, potentially being indistinguishable from the signal spectrum by the minimal scenario.

Having similar philosophy and positing the DM interpretation, we here propose a {\it new} mechanism to develop a {\it narrow} continuum energy spectrum which would {\it fake} a sharp spike.
The research program to explain the excesses in cosmic ray energy spectra with the ``{\it energy-peak}'' emerging under non-minimal DM frameworks has been initiated by Ref.~\cite{Kim:2015usa}, in which various observations of the energy-peak made in the context of collider physics~\cite{Agashe:2012bn,Agashe:2012fs,Agashe:2013eba,Chen:2014oha,Agashe:2015wwa,massive} have been applied to the GC $\gamma$-ray GeV excess.
As in~\cite{Kim:2015usa}, we begin the discussion with noting that multiple DM species could exist in the Universe, and the relevant DM models constructed upon such a DM framework can bring about not only nontrivial cosmological implications, e.g., ``{\it assisted freeze-out}''~\cite{Belanger:2011ww}, but interesting phenomenology, e.g., ``{\it boosted DM}''~\cite{Agashe:2014yua,Berger:2014sqa,Kong:2014mia}.
In this context, the proposed mechanism involves a non-minimal dark sector containing multiple DM particles.

\begin{figure}[t]
\centering
\includegraphics[scale=0.65]{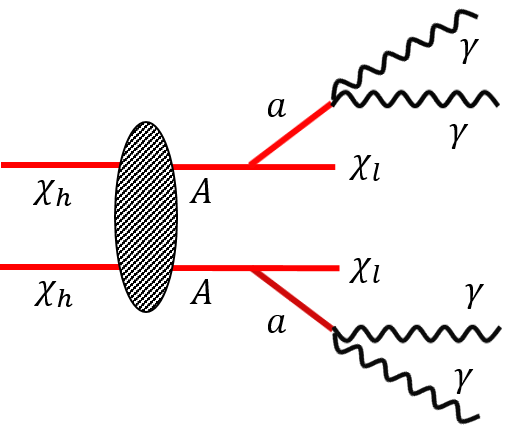}
\caption{\label{fig:model} The dark matter scenario under consideration.}
\end{figure}

For the purpose of simplicity, we introduce two DM species one of which is assumed heavier than the other, denoting henceforth the former and the latter as $\chi_h$ and $\chi_l$, respectively.
The heavier DM communicates to the SM sector not directly but through the lighter DM.
In addition, the heavier DM pair-annihilates into a pair of on-shell intermediate states (denoted as $A$) each of which subsequently decays into the lighter DM together with a dark pion or an axion-like particle (ALP) (denoted as $a$) emitting a couple of photons in the final state.\footnote{In general, $A$ and $\chi_l$ can be either dark or SM sector particles (and may be even unstable) unless they are stringently constrained by other observations.
However, for the sake of simplicity, we assume that they are dark sector particles.}
FIG.~\ref{fig:model} schematizes the DM scenario that we consider throughout this paper.\footnote{The main idea in this paper can be readily applied to the decaying DM scenario, but we keep the annihilating one as a concrete example.}
We point out that although we employ the photon final state as a concrete example for elaborating our mechanism,
it is straightforwardly extensible to other visible particle final states, e.g., $e^+e^-$.

In this DM scenario, the on-shell intermediate particle $A$ comes with a fixed boost factor, leading to a rectangular energy spectrum, hence a rectangular boost distribution for particle $a$.
Due to variations in the boost factor of $a$ the emitted photons develop a {\it continuum} energy spectrum whose width is determined by the mass parameters involved in the entire process demonstrated in FIG.~\ref{fig:model}.
We remark that the relevant scenario invokes from the (relatively) narrow energy spectrum to the broadly-distributed energy spectrum, depending on details of the DM models of interest.
We shall briefly discuss the aspect of the wide continuum bump signature in Sec.~\ref{sec:spectrum}, while primarily focusing on the ``peak-faking'' interpretation and the associated mass spectrum.

Not surprisingly, the mass spectrum of $m_{\chi_h} \gtrsim m_A \gtrsim m_a + m_{\chi_l}$ renders a relatively narrow photon energy spectrum.
As a characteristic feature, the resulting differential energy spectrum is {\it symmetric} in the logarithmic scale, and remarkably, its center position is identified as half the mass of particle $a$, $m_a/2$~\cite{Agashe:2012bn,1971NASSP.249.....S}.
Therefore, these structural properties enable us to not only distinguish this DM model scenario from other standard DM interpretations, but probe/measure the mass parameters of some dark sector particles.

To present our main idea, this paper is organized as follows.
We first begin with the DM model under consideration in the next section.
In Sec.~\ref{sec:spectrum}, we discuss the energy spectra of relevant visible particles arising from the DM model introduced in Sec.~\ref{sec:model}, mainly focusing on their functional structure.
We then apply the main idea to a couple of photon peaks in Sec.~\ref{sec:applications}: i) Fermi-LAT 130 GeV excess and ii) 3.5 keV excess.
Sec.~\ref{sec:conclusions} is reserved for our conclusions.

\section{Dark matter model \label{sec:model}}

We here discuss a class of DM models for which our main idea is applied, and offer a viable DM model that realizes the relevant scenario.
As briefly mentioned earlier, we imagine that the dark sector is non-minimal, meaning that there exist more than one DM candidate.
Although the arbitrary number of DM species could be introduced, we employ only two types of DM particles, the heavier ($\chi_h$) and the lighter ($\chi_l$), for simplicity.

The lighter DM is assumed to directly communicate to the SM sector, whereas the heavier DM is set out to have interactions with the SM sector via the lighter DM.
In this sense, the relic abundance of $\chi_h$ can be computed by the scheme of ``{\it assisted freeze-out}''~\cite{Belanger:2011ww}.
We further assume that $\chi_h$ has a contact with $\chi_l$ {\it not} directly, {\it but} via an on-shell intermediate state $A$, i.e., a pair of heavier DM particles annihilate into a pair of $A$'s each of which subsequently decays into $\chi_l$ together with a dark pion or an ALP as depicted in FIG.~\ref{fig:model}.
Finally, the dark pion or the ALP decays into a photon pair whose energy spectrum is the major interest of this paper.
We remark that it is not a necessary condition for particle $A$ to get pair-produced, i.e., $A$ could be produced in association with another particle $A'$ as the detailed dynamics of $A'$ is irrelevant to the later argument and formalism.
Similarly, $A$ does not need to directly decay into the lighter DM, i.e., $\chi_l$ could be replaced by other heavy particles whose detailed dynamics does not affect the later argument and formalism.
In this context, the model set-up demonstrated in FIG.~\ref{fig:model} is a simplified version.

A possible realization of the DM scenario at hand can be summarized as follows.
We consider two fermionic DM species, $\chi_h$ and $\chi_l$, with an intermediate fermionic state $\psi_A$ and a singlet pseudo-scalar $a$ (e.g., a dark pion or an ALP as mentioned earlier).
Then, the effective Lagrangian required for the DM scenario exhibited in FIG.~\ref{fig:model} is simply described by the following operators:
\bea\label{eq:Model}
\mathcal{L}_{\rm DM} \supset \frac{1}{\Lambda^2}\, \overline{\chi}_h \chi_h \overline{\psi}_A \psi_A + \lambda\, a \overline{\psi}_A\gamma^5\chi_l +\frac{1}{f_a}\, a F_{\mu\nu}\widetilde{F}^{\mu\nu}, \label{eq:effL}
\eea
where $\widetilde{F}^{\mu\nu}$ denotes the dual field strength tensor as usual, and $\Lambda$ and $f_a$ describe the associated suppression scales whose details can be revealed by appropriate UV completion.
The first term ensures an $s$-wave annihilation of the heavier DM, i.e., $\overline{\chi}_h\chi_h \to \psi_A\overline{\psi}_A$, the second induces the decay of $\psi_A$ into $\chi_l$ and $a$, and the last corresponds to two photon decay of $a$ as in FIG.~\ref{fig:model}, respectively.
The stability of $\chi_h$ and $\chi_l$ can be easily achieved with separate symmetries, e.g., U(1)$'\otimes {\rm U(1)}''$~\cite{Belanger:2011ww} or  $Z_2'\otimes Z_2''$~\cite{Agashe:2014yua}.
We again stress that the Lagrangian in Eq.~(\ref{eq:effL}) is a simple realization and there exist a host of other possibilities to accommodate the event topology in FIG.~\ref{fig:model}.
Exhausting all of them is, however, beyond the scope of this paper.

\section{Energy spectrum \label{sec:spectrum}}

In this section, we derive the analytic expression for the gamma-ray energy spectrum and discuss its properties  which could be distinguished from other (standard) scenarios, given the scenario in FIG.~\ref{fig:model}.

\subsection{Derivation of the analytic expression}
Assuming that the heavier DM particles are non-relativistic, their pair annihilation into two $A$'s leads to a fixed boost of particle $A$ (denoted as $\gamma_A$) relating the two mass parameters by $\gamma_A =m_{\chi_h}/m_A$ with $m_i$ symbolizing the mass of particle species $i$.
Since $A$ obtains a non-zero boost factor, the energy of particle $a$ is not monochromatic, but given by a broad spectrum.
The $a$ energy, $E_a$, measured in the laboratory frame is parameterized as
\bea
E_a =E_a^* \left(\gamma_A +\frac{p_a^*}{E_a^*}\sqrt{\gamma_A^2-1}\cos\theta_a^*\right), \label{eq:Ea}
\eea
where $\theta_a^*$ is the emission angle of $a$ in the $A$ rest frame with respect to the boost direction of $A$ and $E_a^*$ is the fixed $a$ energy measured in the rest frame of particle $A$:
\bea
E_a^*=\frac{m_A^2-m_{\chi_l}^2+m_a^2}{2m_A}.
\eea
If $A$ is either a scalar or produced in an {\it unpolarized} way, then $\cos\theta_a^*$ becomes a flat variable, resulting in a rectangular distribution in $E_a$ by a simple chain rule whose range is given by
\bea
E_a \in \left[E_a^*\gamma_A-p_a^*\sqrt{\gamma_A^2-1},\;E_a^*\gamma_A+p_a^*\sqrt{\gamma_A^2-1} \right]. \label{eq:Eadist}
\eea

Similarly to Eq.~(\ref{eq:Ea}), the observed photon energy for a fixed $\gamma_a$ is expressed as
\bea
E_{\gamma}=E_{\gamma}^*\left(\gamma_a+\sqrt{\gamma_a^2-1}\cos\theta_{\gamma}^*\right),\label{eq:Egamma}
\eea
where $\theta_{\gamma}^*$ denotes the intersecting angle between its emission direction and the boost direction of particle $a$ in the $a$ rest frame and $E_{\gamma}^*$ is the fixed photon energy measured in the $a$ rest frame, that is, half the mass of particle $a$:
\bea
E_{\gamma}^*=\frac{m_a}{2}.
\eea
Unlike $\gamma_A$ in the case of particle $a$, $\gamma_a$ is not single-valued but distributed.
Denoting its distribution by $g(\gamma_a)$, from Eq.~(\ref{eq:Eadist}) we find the unit-normalized expression (or equivalently, probability distribution function) for $g(\gamma_a)$ as
\bea
g(\gamma_a)=\frac{m_a}{2p_a^*\sqrt{\gamma_A^2-1}}\Theta(\gamma_a-\gamma_a^-)\Theta(\gamma_a^+-\gamma_a),
\eea
where $\Theta(x)$ is the usual Heaviside step function, and $\gamma_a^{\pm}$ are defined by
\bea
\gamma_a^{\pm}\equiv \frac{E_a^*}{m_a}\gamma_A\pm\frac{p_a^*}{m_a}\sqrt{\gamma_A^2-1}.\label{eq:gammaapm}
\eea
Here we used the fact that $\cos\theta_{\gamma}^*$ is a flat variable so that $g(\gamma_a)$ develops a rectangular distribution as well.
We then find that for any fixed $\gamma_a$, the unit-normalized differential energy distribution is
\bea
\left.\frac{1}{\Gamma}\frac{d\Gamma}{dE_{\gamma}}\right|_{{\rm fixed }\gamma_a}=\frac{1}{2E_{\gamma}^*\sqrt{\gamma_a^2-1}}\Theta(E_{\gamma}-E_{\gamma}^-)\Theta(E_{\gamma}^+-E_{\gamma}),\label{eq:Egammadistfixgamma}
\eea
where $E_{\gamma}^{\pm}$ can be obtained by setting $\cos\theta_{\gamma}^*$ to be $\pm1$ in Eq.~(\ref{eq:Egamma}):
\bea
E_{\gamma}^{\pm}\equiv E_{\gamma}^*(\gamma_a\pm\sqrt{\gamma_a^2-1}). \label{eq:EgammaRange}
\eea
Denoted by $f(E_{\gamma})$, the expression for the unit-normalized total energy spectrum can be obtained by summing Eq.~(\ref{eq:Egammadistfixgamma}) over all relevant $\gamma_a$'s, that is,
\bea
f(E_{\gamma})&=&\int^{\gamma_a^{\max}}_{\gamma_a^{\min}}d\gamma_a \frac{g(\gamma_a)}{2E_{\gamma}^*\sqrt{\gamma_a^2-1}} \label{eq:integralrep}\\
&=&\frac{m_a}{4E_{\gamma^*}p_a^*\sqrt{\gamma_A^2-1}}\left\{\log\left[\sqrt{(\gamma_a^{\max})^2-1}+\gamma_a^{\max}\right] \right. \nonumber \\
&&\left. -\log\left[\sqrt{(\gamma_a^{\min})^2-1}+\gamma_a^{\min}\right] \right\}, \label{eq:analexpE}
\eea
where $\gamma_a^{\min}$ and $\gamma_a^{\max}$ are defined as
\bea
\gamma_a^{\min}\equiv \max\left[\gamma_a^-,\frac{1}{2}\left(\frac{E_{\gamma}}{E_{\gamma}^*}+\frac{E_{\gamma}^*}{E_{\gamma}} \right) \right], \;\gamma_a^{\max}\equiv \gamma_a^+. \label{eq:gammaadef}
\eea
Since $g(\gamma_a)$ is upper-bounded, $\gamma_a^{+}$ determines the spanning range of $f(E_{\gamma})$ as follows:
\bea
\frac{E_{\gamma}}{E_{\gamma}^*}\in \left[\gamma_a^{+}-\sqrt{(\gamma_a^{+})^2-1},\; \gamma_a^{+}+\sqrt{(\gamma_a^{+})^2-1}\right].
\eea
We finally remark that in the actual data analysis with concrete examples, all prefactors in Eq.~(\ref{eq:analexpE}) are eventually absorbed into the overall normalization parameter $N$, and as a consequence the shape of $f(E_{\gamma})$ is completely determined by $\gamma_a^+$, $\gamma_a^-$, and $E_{\gamma}^*$, i.e., there are four independent fit parameters.

\subsection{Functional properties and discussions}

\begin{figure*}[t]
\centering
\includegraphics[width=8.5cm]{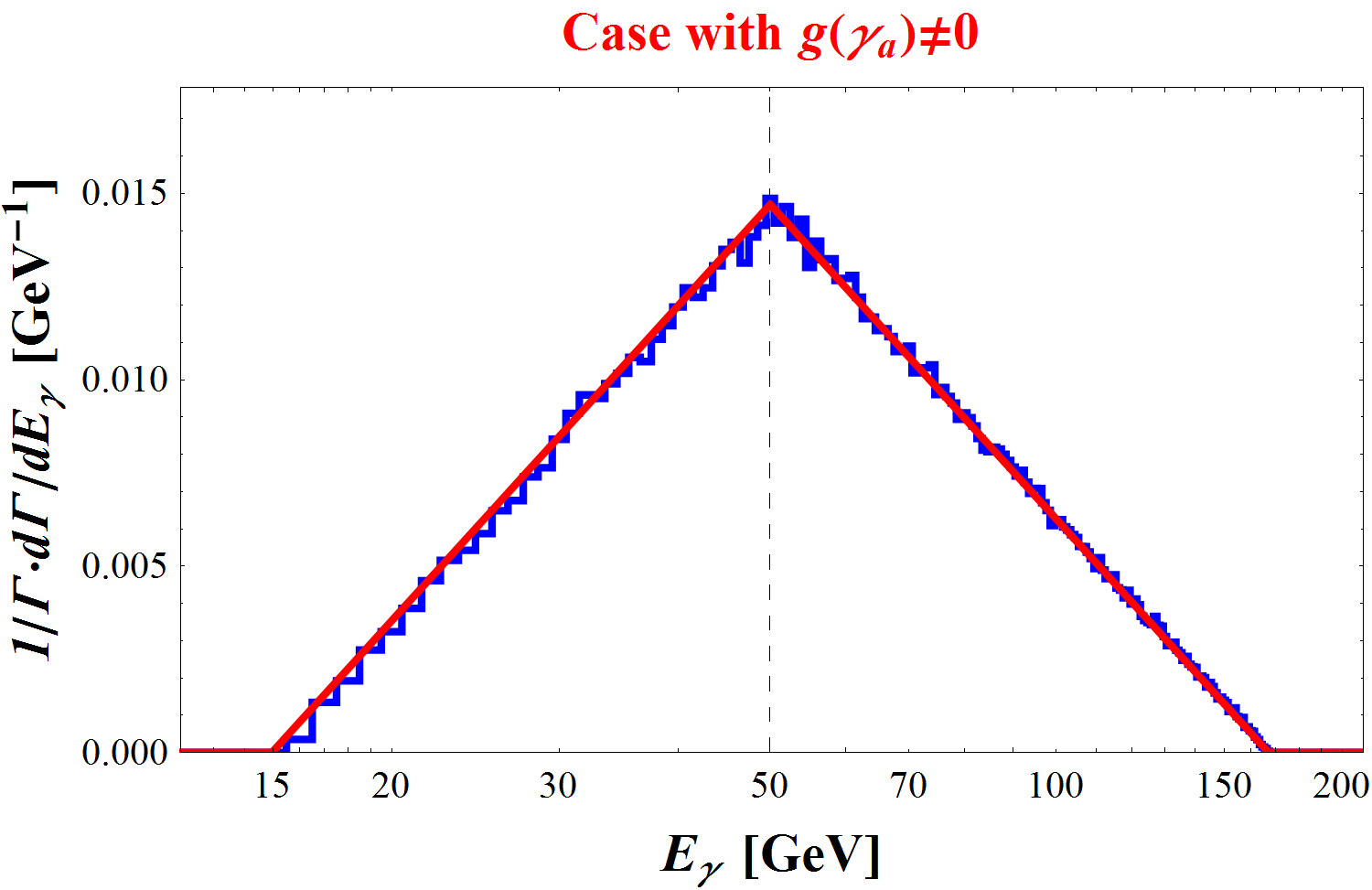}\hspace{0.5cm}
\includegraphics[width=8.5cm]{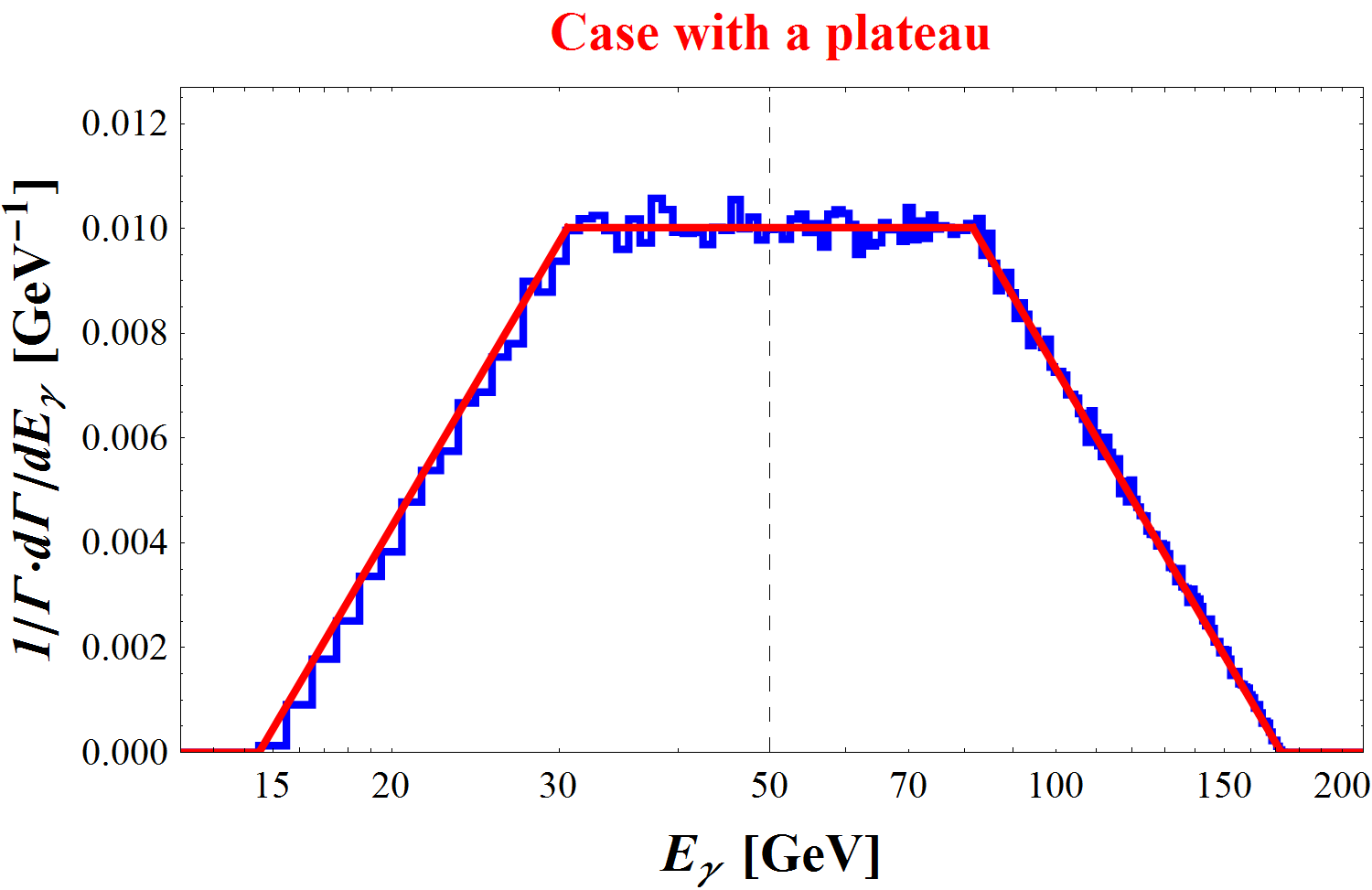}
\caption{\label{fig:theorycurves} Left panel: the gamma-ray energy spectrum with a peak. The chosen mass spectrum is $(m_{\chi_h},\;m_A,\;m_{\chi_l},\;m_a)=(237.5,\; 200,\;50,\;100)$ GeV. The simulated data and corresponding theory expectation are represented by the blue histogram and red line, respectively. Right panel: the gamma-ray energy spectrum with a plateau. The chosen mass spectrum is the same as in the left panel with $m_A$ replaced by 170 GeV.   }
\end{figure*}

To discuss functional properties of the photon energy spectrum, we first revisit the expression of $E_{\gamma}$ for a fixed $\gamma_a$ shown in Eq.~(\ref{eq:Egamma}).
Since $\cos\theta_{\gamma}^*$ spans $-1$ to $+1$, the range of $E_{\gamma}$ is trivially given by
\bea
\frac{E_{\gamma}}{E_{\gamma}^*}\in \left[\gamma_a-\sqrt{\gamma_a^2-1},\; \gamma_a+\sqrt{\gamma_a^2-1}\right],
\eea
as also expressed in Eq.~(\ref{eq:EgammaRange}).
One remarkable feature from the above range is the fact that the lower (upper) end in the right-hand side is smaller (greater) than 1, implying that $E_{\gamma}^*$ is the only commonly-included energy value for {\it any} $\gamma_a$.
Moreover, we observe that $E_{\gamma}^*$ is the geometric mean of minimum and maximum energy values, i.e., $(E_{\gamma}^*)^2=E_{\gamma}^{\min}E_{\gamma}^{\max}$.
This again implies that the $E_{\gamma}^*$ value is located at the center of the $E_{\gamma}$ distribution for a given $\gamma_a$ in the {\it logarithmic} scale.

As mentioned before, the shape of $E_{\gamma}$ distribution for a fixed $\gamma_a$ is rectangular due to the flatness of $\cos\theta_{\gamma}^*$ in Eq.~(\ref{eq:Egamma}).
In order to obtain the overall energy distribution, one should ``stack up'' all of such rectangles contributing to the energy distribution.
This statement is already formulated in Eq.~(\ref{eq:integralrep}) as a Lebesque-type integral representation.
Hence, one would expect that the resultant photon energy distribution contains a unique peak at $E_{\gamma}=E_{\gamma}^*$ in conjunction with the observation made in the previous paragraph.
Indeed, there arises a subtlety here: the validity of this expectation depends whether or not the smallest boost factor of particle $a$ approaches 1.

The condition for $\gamma_a=1$ (i.e., solving Eq.~(\ref{eq:gammaapm}) with $\gamma_a^-$ being 1) is $E_a^*=m_a\gamma_A$, representing a hypersurface formed by $m_A$, $m_{\chi_h}$, and $m_{\chi_l}$ for any fixed $m_a$, i.e., only delicately chosen mass spectra can attain this condition.
Once $\gamma_a=1$ is available, clearly, $E_{\gamma}^*$ appears as a unique and cusp-structured peak~\cite{Agashe:2012bn,Chen:2014oha}.
The left panel in FIG.~\ref{fig:theorycurves} demonstrates an example spectrum in this category in the logarithmic scale.
The chosen mass spectrum is $(m_{\chi_h},\;m_A,\;m_{\chi_l},\;m_a)=(237.5,\; 200,\;50,\;100)$ GeV, and events were generated by pure phase space.
The simulated events are binned to the blue histogram, and the associated theory prediction is shown by the red line.
We clearly see that the theory expectation can reproduce the data well enough, and the spectrum is {\it symmetric} with respect to $E_{\gamma}^*=m_a/2=50$ GeV (indicated by a black dashed line) in this scale.
We also remark that both sides of the distribution look like straight lines, which can be easily seen from Eq.~(\ref{eq:analexpE}) with $\gamma_a^{\min}=\frac{1}{2}\left( \frac{E_{\gamma}}{E_{\gamma}^*}+\frac{E_{\gamma}^*}{E_{\gamma}} \right)$ in logarithmic $E_{\gamma}$, so that the whole spectrum appears as an isosceles triangle.

On the other hand, $g(\gamma_a)$ starts from the value away from $\gamma_a=1$, {\it no} peak is developed in the middle of the energy distribution.
Instead, a plateau region emerges because even the narrowest rectangle corresponding to the smallest $\gamma_a$ has a finite-sized width.
Nevertheless, it is straightforward that the center of the relevant photon energy spectrum can be identified as $E_{\gamma}^*$ in the logarithmic base.
These expectations are manifestly shown in the right panel of FIG.~\ref{fig:theorycurves}.
The relevant mass spectrum is the same as the previous case with $m_A$ replaced by 170 GeV.
Also, the existence of a plateau region makes the whole energy spectrum appears as an isosceles trapezoid in the logarithmic base.
This plateau structure is a distinguished feature from other energy spectra.
However, its presence may be invisible in actual indirect detection experiments due to the issue of energy resolution.
Clearly, if the plateau is smaller than the relevant resolving power, its existence is rarely identifiable so that the relevant spectrum easily fakes a unimodal distribution like the previous case.
Even for the energy spectrum with a sufficiently sizable plateau, its identification may be unavailable with small statistics, i.e., more data accumulation may be required.

As the proposed mechanism, which is denoted as Scenario iii), is used for explaining narrow peaks, it is interesting to compare and contrast it with the other two conventional scenarios enumerated below.
\begin{itemize} \itemsep3pt \parskip0pt \parsep0pt
\item Scenario i): Photons directly come from DM annihilation/decay.
\item Scenario ii): Photons are emitted as a decay product of the on-shell intermediate particle into which DM annihilates/decays.
\end{itemize}
In Scenario i), the width of the peak is typically caused by the intrinsic energy resolution of detectors, and thus the final energy spectrum is symmetric about the nominal peak.
Even in the logarithmic scale, the spectrum is described by a smooth curve, while it is no more symmetric-looking about the peak.
In Scenario ii), the width of the energy spectrum is physical.
However, identifying the peak position is ambiguous due to the box-like spectral behavior in both the linear and the logarithmic scales.
Therefore, defining the symmetry property of the shape is not available.
The comparisons thus far are summarized in Table~\ref{tab:comparison}.
One can easily see that their respective morphological features differ from one another, and therefore, one is able to pin down the underlying DM scenario with a reasonable amount of signal statistics.

\begin{table}[t]
\centering
\begin{tabular}{c|c|c|c}
\hline \hline
& Scenario i) & Scenario ii) & Scenario iii) \\
\hline
Peak existence & Always & Absent & Sometimes  \\
Plateau existence & Absent & Always & Sometimes \\
Width & Instrumental & Physical & Physical \\
Symmetry in $E$ & Symmetric & Not available & Asymmetric \\
Symmetry in $\log E$ & Asymmetric & Not available & Symmetric \\
Shape in $E$ & Curved & Rectangular & Curved \\
Shape in $\log E$ & Curved & Rectangular & Oblique \\
\hline \hline
\end{tabular}
\caption{\label{tab:comparison} Comparisons of structural properties in the energy spectrum among three DM scenarios defined in the text. Symmetry properties are defined with respect to the relevant peak if available.}
\end{table}

Before closing the current section, we briefly discuss the case of wide continuum bump spectra.
Certainly, such types of spectra arise within a broad realm of relevant parameter space so that the theory prediction in Eq.~(\ref{eq:analexpE}) can be employed to explain continuum cosmic ray excesses.
Due to the unique morphological features discussed so far, the relevant signal spectrum could be easily distinguished from other continuum bumps, taken as strong evidence of a non-trivial dark sector.

\section{Applications \label{sec:applications}}

Armed with the argument in the previous section, we apply the basic idea to a couple of existent cosmic ray peaks: i) 130 GeV line~\cite{Bringmann:2012vr,Weniger:2012tx} and ii) 3.5 keV line~\cite{Bulbul:2014sua,Boyarsky:2014jta}.
Although claiming that these examples could be understood by the proposed mechanism, we admit that the underlying DM models for them might not fall into the scenario depicted in FIG.~\ref{fig:model}.
Nevertheless, we emphasize that the applicability of the relevant technique is restricted to neither these two examples nor gamma-ray spectra.
In other words, nothing precludes us from applying the DM interpretation at hand for any of future excessive cosmic ray signals with a narrow peak.

\subsection{130 GeV line}

Our first example is the famous 130 GeV line whose original data was collected by the FERMI-LAT collaboration.
We basically conduct the fit to the spectrum of the observed 130 GeV gamma-ray excess with the expected shape in Eq.~(\ref{eq:analexpE}) with the overall normalization parameter $N_S$ being added:
\bea
f_S(E_{\gamma})&=&N_S\left\{\log\left[\sqrt{(\gamma_a^{\max})^2-1}+\gamma_a^{\max}\right] \right. \nonumber \\
&&\left. -\log\left[\sqrt{(\gamma_a^{\min})^2-1}+\gamma_a^{\min}\right] \right\} . \label{eq:sigtemp}
\eea
The relevant data points and errors are taken from Reg4 with the ULTRACLEAN event class in Ref.~\cite{Weniger:2012tx}.
Since the measured bin counts contain the contributions from backgrounds, the relevant fit is performed simultaneously with the background template.
We assume that the backgrounds can be parameterized by a simple (gradually-falling) power law such as
\bea
f_B(E_{\gamma})=N_B\left( \frac{E_{\gamma}}{E_{\gamma}^*} \right)^{-p}, \label{eq:bgtemp}
\eea
where $N_B$ is the normalization parameter for backgrounds and $p$ encodes the background shape.
Here $E_{\gamma}^*$ is the same $E_{\gamma}^*$ contained in $\gamma_a^{\min}$ of Eq.~(\ref{eq:sigtemp}).
Therefore, the entire data set is fitted with the combination of $f_S(E_{\gamma})$ in Eq.~(\ref{eq:sigtemp}) and $f_B(E_{\gamma})$ in Eq.~(\ref{eq:bgtemp}), i.e.,
\bea
f_{\rm total}(E_{\gamma})=f_S(E_{\gamma})+f_B(E_{\gamma}). \label{eq:tottemp}
\eea

\begin{figure*}[t]
\centering
\includegraphics[width=8.1cm]{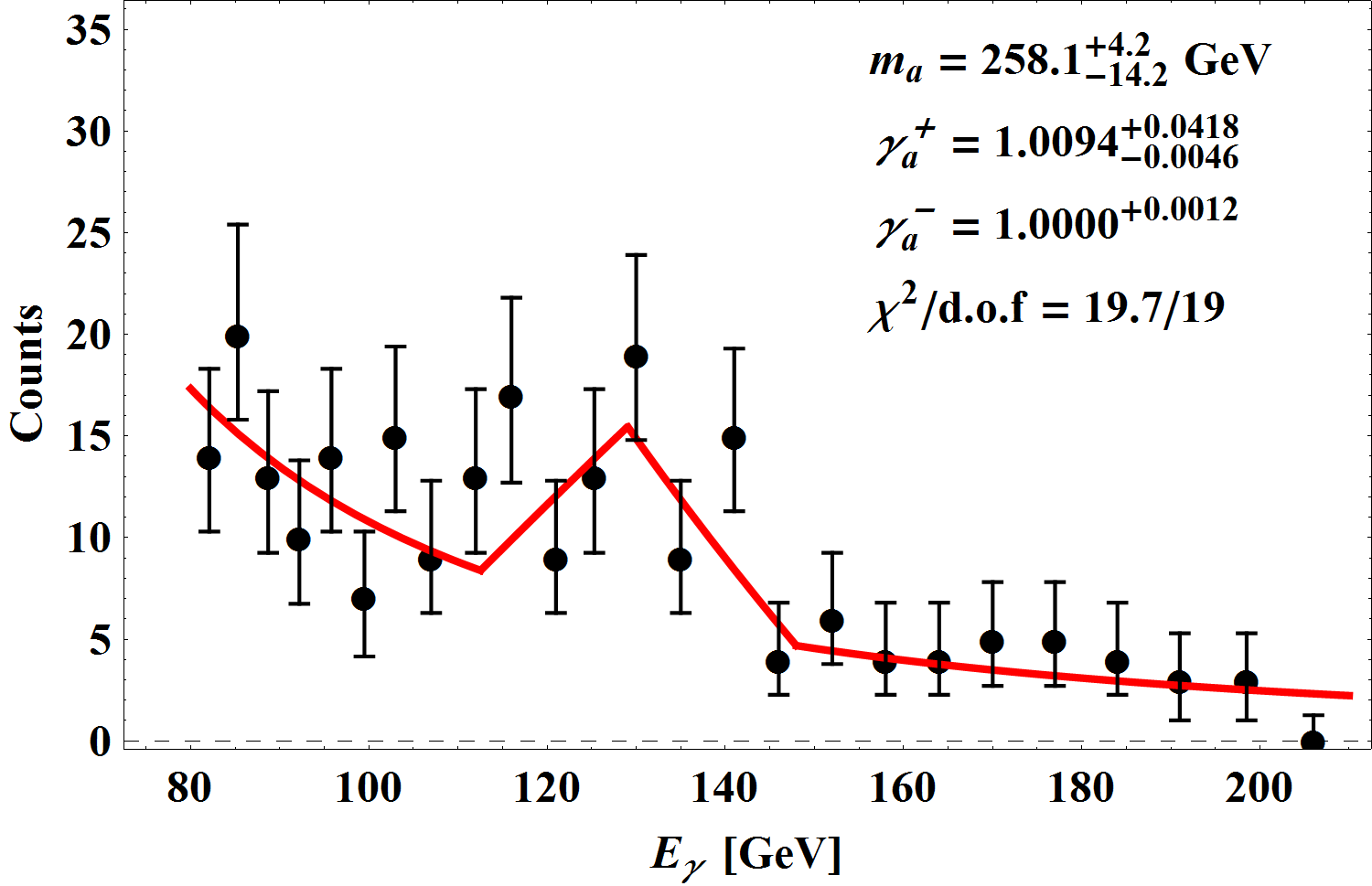} \hspace{0.6cm}
\includegraphics[width=8.1cm]{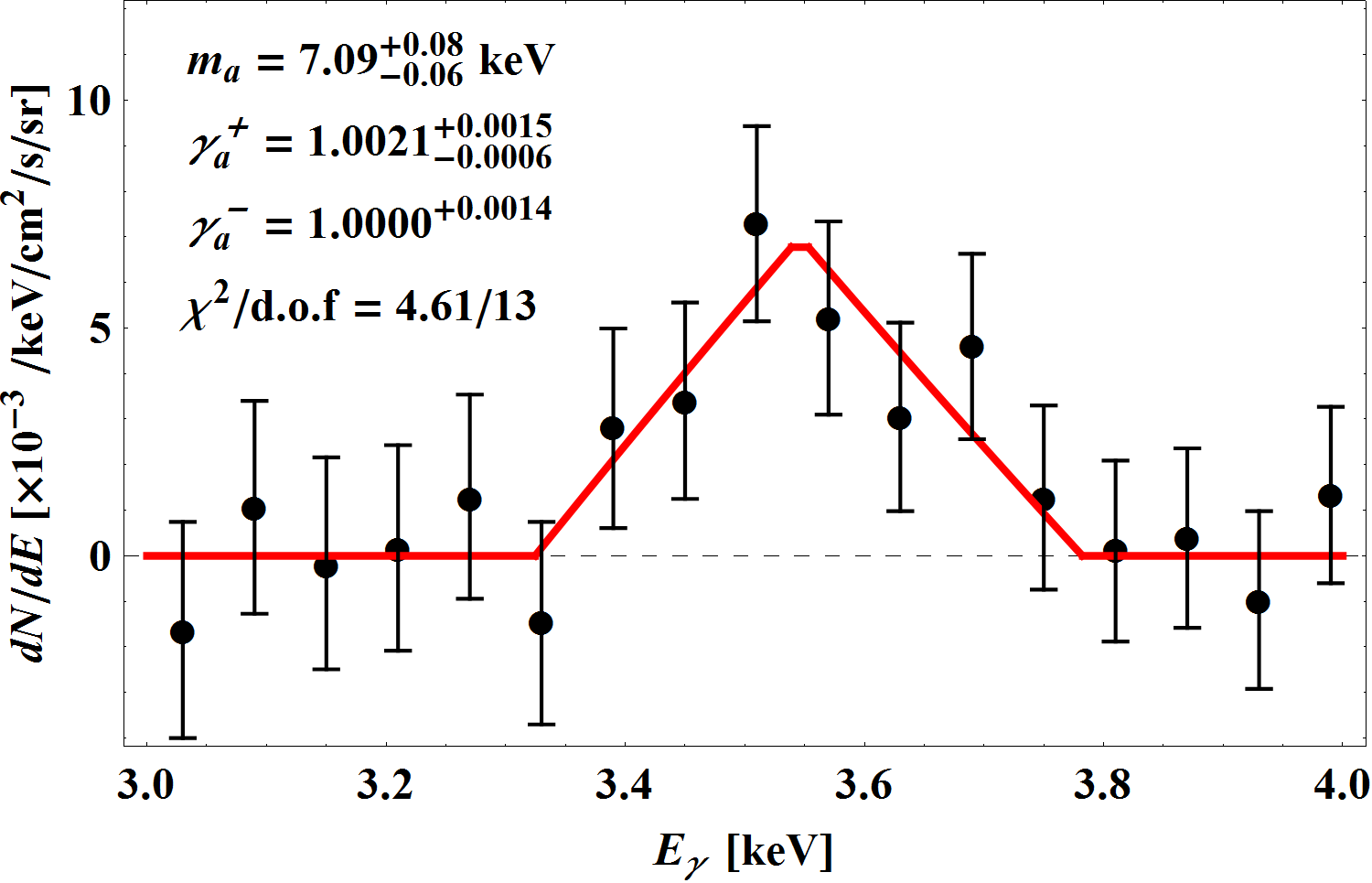} \\ \vspace{0.3cm}
\includegraphics[width=8.2cm]{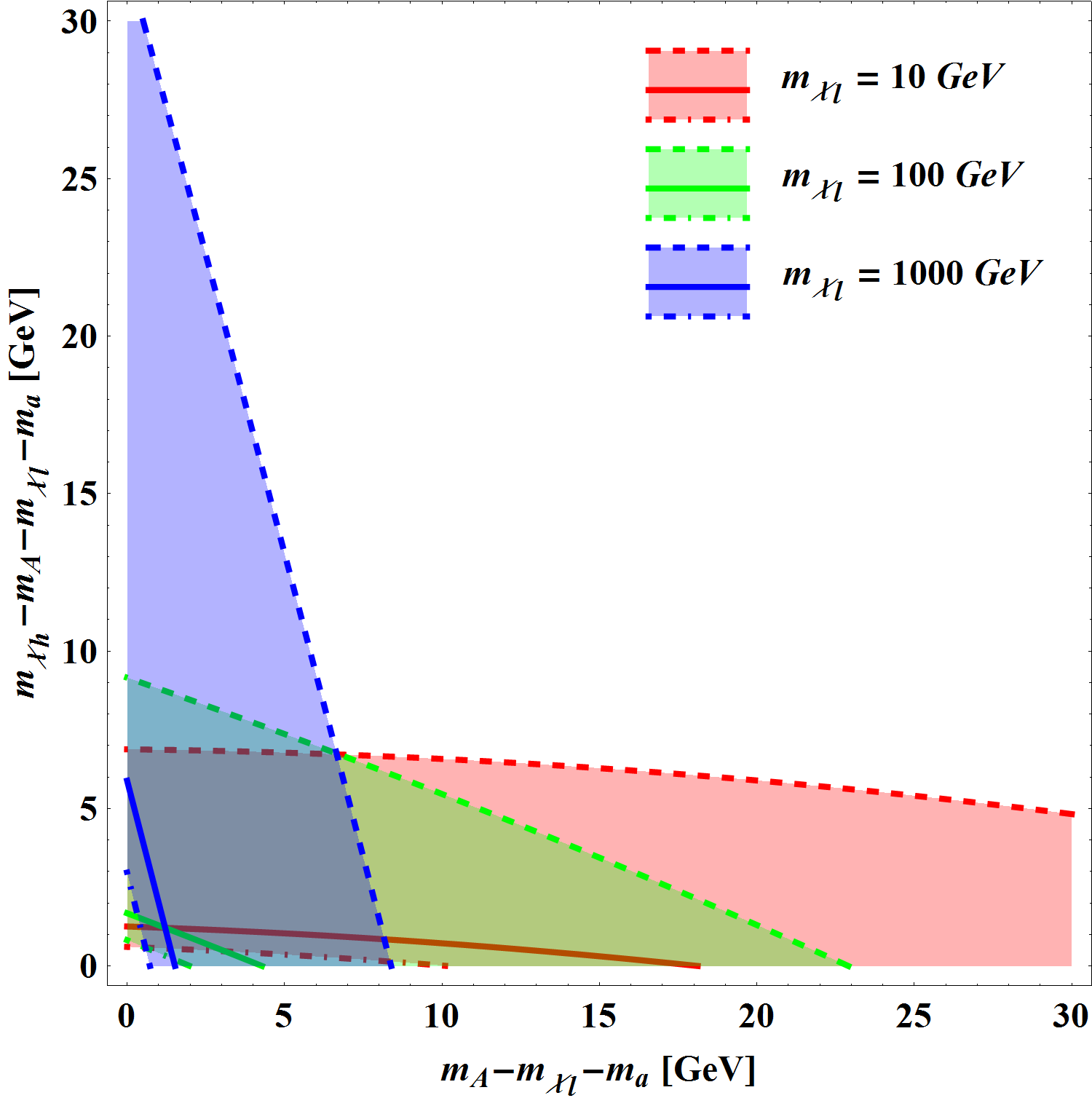} \hspace{0.4cm}
\includegraphics[width=8.5cm]{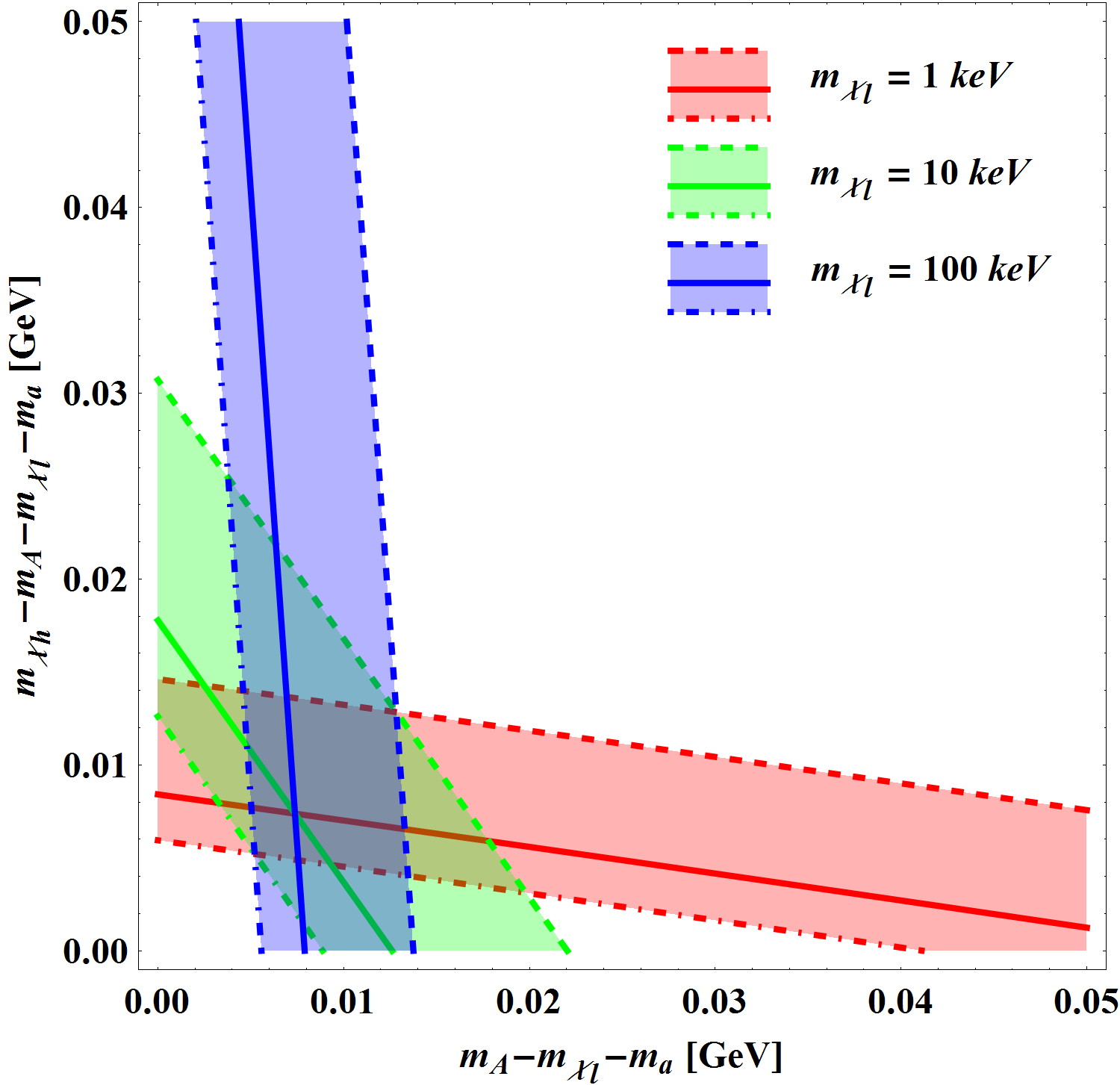}
\caption{\label{fig:fitresults} Upper-left panel: the 130 GeV $\gamma$-ray energy spectrum of the DM and background components taken from Ref.~\cite{Weniger:2012tx}. The fit is performed with 25 data points (black dots), and the best-fitted is represented by the red curve. Upper-right panel: the 3.5 keV X-ray energy spectrum of the DM components taken from Ref.~\cite{Boyarsky:2014jta}. The fit is performed with 17 data points. Lower-left panel: the allowed mass space for the 130 GeV line along with 1$\sigma$ variations of $m_a$ and $\gamma_a^+$ for three different masses of particle $\chi_l$. Lower-right panel: the allowed mass space for the 3.5 keV line.}
\end{figure*}

Our fit result is shown in the upper-left panel of FIG.~\ref{fig:fitresults}.
The data points and associated error bars are represented by black dots and black lines, correspondingly, while the red solid curve shows the best-fitted model.
We clearly see that our fit is in a rather good agreement with the gamma-ray energy spectrum, i.e., the template given in Eq.~(\ref{eq:tottemp}) reproduces the data sufficiently well.
More quantitatively speaking, we find that the $\chi^2$ value is 19.7 for 19 degrees of freedom (i.e., 25 data points subtracted by 6 fitting parameters such as $N_S$, $N_B$, $\gamma_a^+$, $\gamma_a^-$, $E_{\gamma}^*$, and $p$) between 80 and 210 GeV.
This number is quite comparable to that from conventional scenarios, suggesting that our DM scenario be considered (equally) plausible in explaining the observed data.

The fit also tells us useful information. First of all, we extract the mass of particle $a$ from the measurement of $E_{\gamma}^* (=m_a/2)$, that is,
\bea
m_a^{\rm ext} = 258.1^{+ 4.2}_{-14.2} \hbox{ GeV},
\eea
where the errors here are evaluated by 1$\sigma$ statistical uncertainty.
The other mass parameters can be also estimated by the measured $\gamma_a^{+}$ and $\gamma_a^{-}$, for which the best-fitted numbers are reported as
\bea
(\gamma_a^+)^{\rm ext} &=&1.0094^{+0.0418}_{-0.0046}, \\
(\gamma_a^-)^{\rm ext} &=&1.0000^{+0.0012},
\eea
respectively.
Again, the errors are reported by 1$\sigma$ statistical uncertainty.
Note that the lower error for $(\gamma_a^-)^{\rm ext}$ is not provided because there is no sensitivity to the values of $\gamma_a^-<1$ according to the definition of $\gamma_a^{\min}$ in Eq.~(\ref{eq:gammaadef}).
From Eq.~(\ref{eq:gammaapm}), we obtain the expressions for $E_a^*$ and $p_a^*$ in terms of $\gamma_a^{\pm}$
\bea
E_a^*\gamma_A&=&\frac{m_a(\gamma_a^++\gamma_a^-)}{2}, \\
p_a^*\sqrt{\gamma_A^2-1}&=&\frac{m_a(\gamma_a^+-\gamma_a^-)}{2},
\eea
and the difference between the former squared and the latter squared leads the following mass relation:
\bea
\frac{m_{\chi_h}^2}{m_A^2}-1+\frac{m_A^2-m_{\chi_l}^2+m_a^2}{2m_A m_a}=\gamma_a^+\gamma_a^-, \label{eq:contourbasic}
\eea
with which we perform a scan of allowed mass space.

As $\gamma_a^-$ shows the smallest error, we simply fix it to be unity for phenomenological purpose.
Instead of doing three-dimensional scanning, we choose three different $m_{\chi_l}$ values, 10, 100, and 1000 GeV to find the allowed regions in terms of $m_{\chi_h}$ and $m_A$.
The lower-left panel of FIG.~\ref{fig:fitresults} demonstrates the allowed parameter space in the plane of $(m_A-m_{\chi_l}-m_a)$ versus $(m_{\chi_h}-m_A-m_{\chi_l}-m_a)$.
The red, green, and blue regions are for $m_{\chi_l}=10$, 100, and 1000 GeV, respectively.
Solid curves are the contours evaluated with the best-fitted $m_a$ and $\gamma_a^+$.
On the other hand, dashed (dot-dashed) curves are the ones evaluated with $m_a+\delta m_a$ and $\gamma_a^++\delta\gamma_a^+$ ($m_a-\delta m_a$ and $\gamma_a^+-\delta\gamma_a^+$), so that one can get some intuition on the mass spectrum allowed by 1$\sigma$ variations of the relevant parameters.
We observe that the viable mass spectra are more or less compact.
This is not surprising because the narrow peak enforces a degenerate mass spectrum not to obtain too large boost in any of the steps in FIG.~\ref{fig:model}.
Dark sector scenarios featuring such a mass spectrum could be achieved by a symmetry.
Building realistic DM models is, however, beyond the scope of this paper, so we do not further pursue this direction here.

\subsection{3.5 keV line}

Our next example is the well-known 3.5 keV line.
As in the previous case, the fit is performed to the spectrum of the observed 3.5 keV gamma-ray excess with the signal template given in Eq.~(\ref{eq:sigtemp}).
The relevant data points and associated errors are taken from the processed data for the MOS spectrum of the central region of the Andromeda galaxy (M31) found in Ref.~\cite{Boyarsky:2014jta}.
The numbers to be used contain only the DM component, i.e., the background component is already subtracted.
Therefore, we execute the fit procedure only with the signal template, not introducing any background template unlike the previous case.

The fit result is exhibited in the upper-right panel of FIG.~\ref{fig:fitresults}.
Again, the data points and the error bars are represented by black dots and black lines, respectively, while the red solid line describes the best-fitted model.
Our fit is in a very good agreement with the measured energy spectrum.
The reported $\chi^2$ value is 4.61 for 13 degrees of freedom (i.e., 17 data points subtracted by 4 fitting parameters such as $N_S$, $\gamma_a^+$, $\gamma_a^-$, and $E_{\gamma}^*$) between 3 and 4 keV.
This number is significantly improved, compared with that from standard interpretations, and therefore, our DM framework can be considered as a plausible scenario accommodating the observed spectrum.

Speaking of various best-fit parameters, we first find that the extracted mass parameter for particle $a$ is
\bea
m_a^{\rm ext}=7.09^{+0.08}_{-0.06} \hbox{ keV},
\eea
where errors are estimated by 1$\sigma$ statistical uncertainty.
We find the best-fit values for $\gamma_a^+$ and $\gamma_a^-$ are
\bea
(\gamma_a^+)^{\rm ext}&=&1.0021^{+0.0015}_{-0.0006}, \\
(\gamma_a^-)^{\rm ext}&=&1.0000^{+0.0014},
\eea
respectively, together with 1$\sigma$ statistical uncertainty.
In order to obtain the allowed mass space within 1$\sigma$ variations of the relevant parameters, we follow the same procedure in the previous case for three different $m_{\chi_l}$ masses, 1, 10, and 100 keV.
The lower-right panel of FIG.~\ref{fig:fitresults} demonstrates the allowed region again in the plane of $(m_A-m_{\chi_l}-m_a)$ versus $(m_{\chi_h}-m_A-m_{\chi_l}-m_a)$.
The red, green, and blue regions correspond to $m_{\chi_l}=1$, 10, and 100 keV, respectively.

\section{Conclusions \label{sec:conclusions}}

Dark matter indirect detection offers an excellent opportunity to confirm the existence of DM.
Several experimental collaborations have already reported anomalous phenomena in the relevant cosmic ray energy spectrum.
Particular attention has been paid to sharply-peaked signals due to the easiness of DM interpretation.
Typical models assume that (non-relativistic) DM particles directly annihilate or decay into visible ones.
In this minimal DM scenario, the narrow width of the peak is typically understood as the one stemming from the imperfection of cosmic ray detectors.

We have rather taken the viewpoint that this width can be physically induced, and proposed a {\it new} mechanism to realize it with the assumption of a non-minimal dark sector.
Two DM species were introduced, and the heavier one is assumed to annihilate to the on-shell intermediate state which subsequently decays into the lighter one and an unstable particle.
This unstable particle further decays into a pair of visible particles which can be a source of anomalous peaks in the cosmic ray energy spectra.
We showed that the signal spectrum can be narrow enough to fake a sharp spike with a suitable choice of the associated mass parameters.

The shape of the full signal spectrum was derived, and several interesting functional properties were discussed.
We pointed out that the peak position, one of the fit parameters, is immediately identified as half the mass of the above-mentioned unstable particle.
We also showed that other mass parameters can be estimated by other fit parameters.
We then enumerated various morphological features to be utilized for distinguishing several DM scenarios in which the sharp peak signature is available.
The viability of the relevant strategy was assessed with two real observational data sets, 130 GeV line and 3.5 keV line.
We found that both of them can be well-described by the theoretical expectation in our DM scenario, and that each allowed parameter space is fairly large.

Finally, we emphasize that our DM scenario and the associated data analysis are not restricted to photon energy peaks, i.e., any cosmic ray energy peaks can take advantage of them.
We strongly encourage people to pursue the direction exploited in this paper as well whenever cosmic ray peaks are observed.


\section*{Acknowledgments}

We thank Kaustubh Agashe for useful discussions.
D.~K. is supported by the LHC Theory Initiative postdoctoral fellowship (NSF Grant No. PHY-0969510), and J-C.~P. is supported by Basic Science Research Program through the National Research Foundation of Korea funded by the Ministry of Education (NRF-2013R1A1A2061561).
We appreciate CETUP* (Center for Theoretical Underground Physics and Related Areas) for its hospitality enabling us to initiate this work.

\end{document}